\documentclass[nofootinbib,aps,pra,showpacs,superscriptaddress,preprint]{revtex4}
\usepackage{amsmath}
\usepackage{subfig}
\usepackage{graphicx}
\usepackage{lxfonts}

\catcode`ð=\active
\defð{\u{g}}
\catcode`Ð=\active
\defÐ{\u{G}}
\catcode`Ý=\active
\defÝ{\. I}
\catcode`ö=\active
\defö{\"{o}}
\catcode`Ö=\active
\defÖ{\"O}
\catcode`ü=\active
\defü{\"{u}}
\catcode`Ü=\active
\defÜ{\"{U}}
\catcode`Þ=\active
\defÞ{\c{S}}
\catcode`þ=\active
\defþ{\c{s}}
\catcode`ý=\active
\defý{{\i}}
\catcode`ç=\active
\defç{\d{c}}
\catcode`Ç=\active
\defÇ{\d{C}}

\begin{document}

\title{Non-Central Potentials, Exact Solutions and Laplace Transform Approach}
\author{\small Altuð Arda}
\email[E-mail: ]{arda@hacettepe.edu.tr}\affiliation{Department of
Physics Education, Hacettepe University, 06800, Ankara,Turkey}
\author{\small Ramazan Sever}
\email[E-mail: ]{sever@metu.edu.tr}\affiliation{Department of
Physics, Middle East Technical  University, 06531, Ankara,Turkey}

\begin{abstract}
Exact bound state solutions and the corresponding wave functions
of the Schrödinger equation for some non-central potentials
including Makarov potential, modified-Kratzer plus a ring-shaped
potential, double ring-shaped Kratzer potential, modified
non-central potential and ring-shaped non-spherical oscillator
potential are obtained by using the Laplace transform approach.
The energy spectrums of the Hartmann potential, modified-Kratzer
potential and ring-shaped oscillator potential are also briefly
studied as special cases.
It is seen that our analytical results for all these potentials are consistent with those obtained
by other works. We also give some numerical results obtained for the modified non-central potential for different values of the
related quantum numbers.
\end{abstract}

\maketitle

\newpage

\section{Introduction}
An important part of chemistry based on quantum mechanics and also
of nuclear physics include to study ro-vibrational energy levels
of molecules and atoms having multi-electrons, the distorted
nucleus and the correlation states of quantum fluid systems [1].
Describing ring-shaped molecules (like benzene) and interactions
between deformed pairs of nuclei have received many applications
in the above areas of physics [2]. Because of the above
statements, the non-central potentials have been extensively
studied in literature. Moreover, these potentials provide a useful
theoretical ground describing the interaction between the
ring-shaped molecules and the interaction between distorted
nucleus [1, 2].

Non-central potentials including also ring-shaped molecular
potentials include two parts: The spherical harmonic oscillator
potentials and angular dependent potentials as a second part. Such
potentials studying in non-relativistic and/or relativistic
quantum mechanical viewpoints by using different methods could be
listed as: Hartmann [3] and Makarov potential [4] within the
supersymmetric approach [5, 6], algebraic investigation of the
ring-shaped potential [7], a new anharmonic oscillator potential
studying in terms of hypergeometric functions [8], investigation
of a non-central potential parameterizing with $\hat{C}, C, C_{0}$
based on $L^2$-series solution [9], the Coulombic ring-shaped [10]
and Makarov potential [11] within the Nikiforov-Uvarov formalism,
relativistic searching of Makarov potential by factorization
method [12], double ring-shaped oscillator potential within the
supersymmetric formalism [13], the Hartmann potential via Laplace
transforms [14], searching of some non-central potentials by using
exact quantization rule [15], \textit{etc.}.

The non-central potentials are needed to obtain better results
than those given of central potentials about the dynamical
properties of the molecular structures and interactions [16].
Meanwhile, these potentials make it possible to obtain algebraic
exact solutions of the Schrödinger equation (SE). To obtain the
exact solutions of the SE for molecular potentials is one of the
basic problems in quantum physics [17]. In this manner, we search
in the present work the exact bound state solutions of the SE for
some non-central potentials including the ring-shaped
non-spherical oscillator potential, the Makarov potential, the
modified Kratzer plus a ring-shaped potential and the double
ring-shaped Kratzer potential. The Laplace transform approach
(LTA) will be used to find the energy levels and the corresponding
wave functions of the above potentials.

The LTA has been widely used to obtain the exact solutions of
central and non-central potentials in the non-relativistic domain
[14, 18-20]. This approach is also used to find some recursion
relations in terms of step-up and step-down operators for the
harmonic oscillator [21]. The LTA describes a simple way for
obtaining the solutions of the SE by reducing it to a first-order
differential equation meaning that whose solutions may be obtained
easily.

The organization of this letter is as follows. In Section II, the
time-independent SE in spherical coordinates is separated into
radial and angular equations for a particle subjected to a
non-central potential. In Section III, the LTA is applied to the
radial SE to obtain the energy spectrum of the above non-central
potentials and the results are compared with those obtained
before.

\section{Equations in Spherical Coordinates}
Time-independent SE in spherical coordinates is written [22]
\begin{eqnarray}
\left\{\vec{\nabla}^2-MV(r,\theta,\phi)+ME_{n\ell}\right\}\Psi(r,\theta,\phi)=0\,,
\end{eqnarray}
where $M=\frac{2m}{\hbar^2}$, $E_{n\ell}$ is the particle energy
and $V(r,\theta,\phi)$ is the potential field giving
\begin{eqnarray}
V(r,\theta,\phi)=V(r)+\frac{V(\theta)}{r^2}+\frac{V(\phi)}{r^2sin^2\theta}\,.
\end{eqnarray}

Writing the total wave function as
\begin{eqnarray}
\Psi(r,\theta,\phi)=\mathcal{R}(r)\Theta(\theta)(sin\theta)^{-1/2}\Phi(\phi)\,\,\,;\mathcal{R}(r)=\frac{R(r)}{r}\,
\end{eqnarray}
and using the method of separation of variables gives the
following equations [14, 23]
\begin{subequations}
\begin{align}
\left\{\frac{d^2}{d\phi^2}-MV(\phi)+m^2\right\}\Phi(\phi)&=0\,,\\
\left\{\frac{d^2}{d\theta^2}-MV(\theta)-\frac{\xi}{sin^2\theta}+\ell^2\right\}\Theta(\theta)&=0\,\,\,\,;\xi=m^2-\frac{1}{4}\,,\\
\left\{\frac{d^2}{dr^2}-MV(r)-\frac{L}{r^2}+ME_{n\ell}\right\}R(r)&=0\,\,\,\,;L=\ell^2-\frac{1}{4}\,.
\end{align}
\end{subequations}
where $m^2$ and $\ell^2$ are separation constants. Throughout this
paper $V(\phi)$ will be $V(\phi)=0$, then Eq. (4a) becomes
\begin{eqnarray}
\left\{\frac{d^2}{d\phi^2}+m^2\right\}\Phi(\phi)&=0\,,
\end{eqnarray}
and its solution
\begin{eqnarray}
\Phi(\phi)=a_ne^{\pm im\phi}\,\,\,\,,m=0, 1, 2, \ldots
\end{eqnarray}

Now we use the LTA applied to Eq. (4c) to find the bound state
solutions and the corresponding wave functions of the non-central
potentials. It is well known that the contributions coming from
the angular part of the potential are placed in the parameter
$\ell$ in Eq. (4b). So, these contributions are taken from related
literature while we are looking for the solutions of Eq. (4c).
\section{Bound State Solutions}
\subsection{Makarov Potential}
Inserting the Makarov potential [4]
\begin{eqnarray}
V(r,
\theta)=\frac{\alpha}{r}+\frac{\beta}{r^2sin^2\theta}+\frac{\gamma
cos\theta }{r^2sin^2\theta}\,,
\end{eqnarray}
into Eq. (4b) we obtain the polar angle equation
\begin{eqnarray}
\left\{\frac{d^2}{d\theta^2}-\frac{1}{sin^2\theta}\,\left(\xi+\frac{M}{r^2}\,\left(\beta+\gamma
cos\theta \right)\right)+\ell^2\right\}\Theta(\theta)=0\,,
\end{eqnarray}
where the parameter $\ell$ is obtained as
$\ell=\left\{\frac{1}{2}\,\left[\beta+m^2+\sqrt{(\beta+m^2)^2-\gamma^2\,}\right]\right\}^{1/2}+s+\frac{1}{2}$
in Ref. [15] and $s$ is a positive integer.

The radial equation (4c) becomes for the Makarov potential
\begin{eqnarray}
\left\{\frac{d^2}{dr^2}-\frac{M\alpha}{r}-\frac{L}{r^2}+ME_{n\ell}\right\}R(r)=0\,,
\end{eqnarray}
Rewriting the radial wave function as
$\psi(r)=r^{-\frac{1}{2}}R(r)$ and inserting it into Eq. (9) we
get
\begin{eqnarray}
\left\{\frac{d^2}{dr^2}+\frac{1}{r}\frac{d}{dr}-\left(\mu^2_{1}+\frac{\mu^2_{2}}{r}+\frac{\mu^2_{3}}{r^2}\right)\right\}\psi(r)=0\,,
\end{eqnarray}
where
\begin{eqnarray}
\mu^2_{1}=-ME_{n\ell}\,\,;\,\,\mu^2_{2}=M\alpha\,\,;\,\,\mu^2_{3}=\frac{1}{4}+L
\end{eqnarray}
In order to obtain an equation suitable for applying the Laplace
transform approach, we take the wave function as
$\psi(r)=r^{\delta}g(r)$ with $\delta$ is a constant in Eq. (10)
and then obtain
\begin{eqnarray}
r^2\frac{d^2g(r)}{dr^2}+(2\delta+1)r\frac{dg(r)}{dr}-\left(\mu^2_{1}r^2+\mu^2_{2}r+\mu^2_{3}-\delta^2\right)g(r)=0\,,
\end{eqnarray}
Physically acceptable wave function has to be finite when $r
\rightarrow \infty$ in Eq. (12), so the parameter $\delta$ should
be $\delta=-\mu_{3}$ and we get
\begin{eqnarray}
\left\{\frac{d^2}{dr^2}-(2\mu_{3}-1)\,\frac{1}{r}\,\frac{d}{dr}-\mu^2_{1}r-\mu^2\right\}g(r)=0\,,
\end{eqnarray}

By applying the Laplace transform defined as [24]
\begin{eqnarray}
\mathcal{L}\left\{g(r)\right\}=f(t)=\int_{0}^{\infty}dr
e^{-tr}g(r)\,,
\end{eqnarray}
to Eq. (13), we obtain
\begin{eqnarray}
\left(t^2-\mu^2_{1}\right)\frac{df(t)}{dt}+\left\{\left(2\mu_{3}+1\right)t+\mu^2_{2}\right\}f(t)=0\,.
\end{eqnarray}
which is a first-order differential equation and its solution is
written as
\begin{eqnarray}
f(t)=a'(t+\mu_{1})^{-(2\mu_{3}+1)}\left(\frac{t+\mu_{1}}{t-\mu_{1}}\right)^{\frac{\mu^2_{2}}{2\mu_{1}}+\frac{2\mu_{3}+1}{2}}\,.
\end{eqnarray}
In order to get single-valued wave functions, it should be
\begin{eqnarray}
\frac{\mu^2_{2}}{2\mu_{1}}+\frac{2\mu_{3}+1}{2}=-n,\,\,\,n=0, 1 ,
2, \ldots
\end{eqnarray}
With the help of this condition, the function in Eq. (16) could be
expended into series as
\begin{eqnarray}
f(t)=a''\sum_{k=0}^{n}\frac{(-1)^{k}n!}{(n-k)!k!}\,(2\mu_{1})^{k}\left(t+\mu_{1}\right)^{-(2\mu_{3}+1)-k}\,,
\end{eqnarray}
where $a''$ is another constant. The solution of Eq. (13) is
immediately obtained by applying the inverse-Laplace
transformation [24]
\begin{eqnarray}
g(r)=a'''r^{2\mu_{3}}e^{-\mu_{1}r}\sum_{k=0}^{n}\frac{(-1)^{k}n!}{(n-k)!k!}\frac{\Gamma(2\mu_{3}+1)}{\Gamma(2\mu_{3}+1+k)}\,(2\mu_{1}
r)^{k}
\end{eqnarray}
which gives finally the eigenfunctions of the Makarov potential
\begin{eqnarray}
R(r)=a_{n}r^{\mu_{3}+\frac{1}{2}}e^{-\mu_{1}r}\,_{1}F_{1}(-n,2\mu_{3}+1,2\mu_{1}r)\,.
\end{eqnarray}
where $a_{n}$ is normalization constant and used the following
properties of hypergeometric functions [25]
\begin{eqnarray}
_{1}F_{1}(-n,\sigma,z)=\sum_{m=0}^{n}\frac{(-1)^{m}n!}{(n-m)!m!}\frac{\Gamma(\sigma)}{\Gamma(\sigma+m)}z^{m}\,.
\end{eqnarray}

Eq. (17) with Eq. (11) gives us the following algebraic expression
of the bound state of the Makarov potential
\begin{eqnarray}
E_{n\ell}=-\,\frac{M\alpha^2}{4\left(n+\frac{1}{2}+\ell\right)^2}\,.
\end{eqnarray}
which is the same result with the ones given in Ref. [15].

Setting in Eq. (7) $\gamma=0$ gives the Hartmann potential and Eq.
(22) turns into
\begin{eqnarray}
E_{n\ell}=-\frac{M\alpha^2}{(2n+\ell'+1)^2}\,\,;\,\,\,\ell'=2\sqrt{\beta+m^2\,}+2s+1\,.
\end{eqnarray}
which is the exact energy states of the Hartmann potential [11,
15].
\subsection{Modified Kratzer plus a Ring-Shaped Potential}
This potential has the form
\begin{eqnarray}
V(r,\theta)=D_{0}\left(1-\frac{r_0}{r}\right)^2+\frac{\beta}{r^2}\,\left(\frac{cos\theta}{sin\theta}\right)^2\,,
\end{eqnarray}
where $D_{0}$ is the association energy and $r_{0}$ is the
equilibrium distance of the molecule. Inserting this into Eqs.
(4b) and (4c), we obtain the angular dependent and radial
equations as
\begin{subequations}
\begin{align}
\left\{\frac{d^2}{d\theta^2}+\beta M-\frac{\xi+\beta M
}{sin^2\theta}+\ell^2\right\}
\Theta(\theta)&=0\,,\\
\left\{\frac{d^2}{dr^2}-\frac{MD_{0}r^2_{0}+L}{r^2}+\frac{2MD_{0}r_{0}}{r}+M(E_{n\ell}-D_{0})\right\}R(r)&=0\,.
\end{align}
\end{subequations}
where the parameter $\ell$ is obtained as
$\ell^2=\left(\sqrt{\beta+m^2\,}+s+1/2\right)^2-\beta$ in Ref.
[15].

We apply the LTA to Eq. (25b) to obtain the exact bound state
solutions and the corresponding eigenfunctions of the modified
Kratzer plus a ring-shaped potential. We follow the same procedure
in the above section, since the radial equations in Eq. (10) and
Eq. (25b) have the same form. In the present case, the parameters
in Eq. (11) become
\begin{eqnarray}
\mu^2_{1}=M(D_{0}-E_{n\ell})\,\,;\,\,\mu^2_{2}=-2MD_{0}r_{0}\,\,;\,\,\mu^2_{3}=L+MD_{0}r^2_{0}+\frac{1}{4}
\end{eqnarray}

Using the same transformations in the above section on the wave
functions and requirements, we obtain the radial eigenfunctions of
the modified Kratzer plus a ring-shaped potential as
\begin{eqnarray}
R(r)=b_{n}r^{\mu_{3}+\frac{1}{2}}e^{-\mu_{1}r}\,_{1}F_{1}(-n,2\mu_{3}+1,2\mu_{1}r)\,,
\end{eqnarray}
where $b_{n}$ is normalization constant.

In order to get a single-valued wave functions the parameters must
be satisfied an equation like Eq. (17) which gives the energy
spectrum of the modified Kratzer plus a ring-shaped potential as
\begin{eqnarray}
E_{n\ell}=D_{0}-\left[\frac{2n+1+\sqrt{MD_{0}r^2_{0}+\ell^2\,}}{2MD_{0}r_{0}}\right]^{-2}\,,
\end{eqnarray}
which is the same result with the one given in Ref. [15]. If
$\beta=0$ in Eq. (24), we have the modified Kratzer potential
whose energy spectrum is obtained from the last equation
\begin{eqnarray}
E_{n\ell}=D_{0}-\left[\frac{2MD_{0}r_{0}}{2n+1+\sqrt{MD_{0}r^2_{0}+\mathcal{L}(\mathcal{L}+1)+\frac{1}{4}\,}}\right]^{2}\,.
\end{eqnarray}
where the parameter $\mathcal{L}$ is defined as
$\mathcal{L}(\mathcal{L}+1)=\ell^2(\beta \rightarrow
0)-\frac{1}{4}$\,. This spectrum is exactly of the modified
Kratzer potential [26].
\subsection{Double Ring-Shaped Kratzer Potential}
This potential is given [27]
\begin{eqnarray}
V(r,\theta)=-2D_{0}\left\{\frac{r_{0}}{r}-\frac{1}{2}\left(\frac{r^2_{0}}{r^2}\right)\right\}
+\frac{1}{r^2}\left(\frac{\beta}{sin^2\theta}+\frac{\gamma}{cos^2\theta}\right)\,,
\end{eqnarray}
which gives for $\beta=\gamma=0$ the Kratzer potential. For this
potential, Eqs. (4b) and (4c) gives the following equations
\begin{subequations}
\begin{align}
\left\{\frac{d^2}{d\theta^2}-\frac{M\gamma}{cos^2\theta}-\frac{\xi+\beta
M }{sin^2\theta}+\ell^2\right\}
\Theta(\theta)&=0\,,\\
\left\{\frac{d^2}{dr^2}+ME_{n\ell}-\frac{2MD_{0}r_{0}}{r}-\frac{MD_{0}r^2_{0}+L}{r^2}\right\}R(r)&=0\,.
\end{align}
\end{subequations}
where the parameter $\ell$ is obtained as
$\ell=\sqrt{\beta^2+m\,}+\sqrt{\frac{1}{4}+\gamma\,}+2s+1$ in Ref.
[15].

Following the same procedure in last two sections we obtain the
energy levels of the double ring-shaped Kratzer potential and the
corresponding eigenfunctions, respectively,
\begin{eqnarray}
E_{n\ell}=-M\left\{\frac{D_{0}r_{0}}{n+\frac{1}{2}+\sqrt{\ell^2+MD_{0}r^2_{0}\,}}\right\}^2\,,
\end{eqnarray}
and
\begin{eqnarray}
R(r)=c_{n}r^{\mu_{3}+\frac{1}{2}}e^{-\mu_{1}r}\,_{1}F_{1}(-n,2\mu_{3}+1,2\mu_{1}r)\,.
\end{eqnarray}
where $c_{n}$ is normalization constant and the parameters in Eq.
(11) have the values in the present case
\begin{eqnarray}
\mu^2_{1}=-ME\,\,;\,\,\mu^2_{2}=2MD_{0}r_{0}\,\,;\,\,\mu^2_{3}=L+MD_{0}r^2_{0}+\frac{1}{4}
\end{eqnarray}
It is seen that the result given in Eq. (32) is consistent with
the one obtained in Ref. [27].
\subsection{Modified Non-Central Potential}
The modified non-central potential is written [28]
\begin{eqnarray}
V(r,\theta)=D\left(1-\frac{a}{r}\right)^2+\frac{\beta}{r^2
sin^2\theta}+\frac{\gamma cos\theta}{r^2 sin^2\theta}\,,
\end{eqnarray}
where the parameter $D$ corresponds to the association energy and
$a$ corresponds to the equilibrium distance of the molecule. The
radial part of this potential is similar to that of the potential
in Eq. (24). By inserting Eq. (35) into Eqs. (4b) and (4c) we
obtain the following angular dependent and radial equations,
respectively,
\begin{subequations}
\begin{align}
\left\{\frac{d^2}{d\theta^2}-\frac{M\beta+\xi+M\gamma cos\theta
}{sin^2\theta}+\ell^2\right\}
\Theta(\theta)&=0\,,\\
\left\{\frac{d^2}{dr^2}+M(E_{n\ell}-D)+\frac{2MDa}{r}-\frac{MDa^2+L}{r^2}\right\}R(r)&=0\,.
\end{align}
\end{subequations}
where the parameter $\ell$ is given as
$\ell=\sqrt{\frac{1}{2}\,}\sqrt{M\beta+m^2+\sqrt{(m^2+M\beta)^2-(M\gamma)^2\,}\,}+s$
in Ref. [16]. In the present case, applying the LTA as in the
above subsections gives the energy spectrum and the corresponding
eigenfunctions of the modified non-central potential
\begin{eqnarray}
E_{n\ell}=D-\left(\frac{2\sqrt{M\,}Da}{2n+1+\sqrt{4MDa^2+4\ell(\ell+1)+1\,}}\right)^2\,,
\end{eqnarray}
and
\begin{eqnarray}
R(r)=d_{n}r^{\mu_{3}+\frac{1}{2}}e^{-\mu_{1}r}\,_{1}F_{1}(-n,2\mu_{3}+1,2\mu_{1}r)\,.
\end{eqnarray}
where $d_{n}$ is normalization constant and
\begin{eqnarray}
\mu^2_{1}=-M(E_{n\ell}-D)\,\,;\,\,\mu^2_{2}=-2MDa\,\,;\,\,\mu^2_{3}=L+MDa^2+\frac{1}{4}\,.
\end{eqnarray}
Now we give the results our numerical analysis for the diatomic
molecule $N_2$ in Table 1 where the energy eigenvalues are given
for different values of quantum numbers and the parameters of
$\beta$ and $\gamma$. Our parameter values for $N_2$ molecule are
as follows: $D=11.9384$ eV, $\mu=7.00335$ amu and $a=1.0940\,\AA$
[26]. From Table 1, we see that the contributions coming from
angular dependent part of the potential in Eq. (37) are negligible
than the results obtained for the pure Kratzer potential (
$\beta=\gamma=0$).
\subsection{Ring-shaped Non-Spherical Oscillator Potential}
As final potential, we search the energy levels and the
corresponding wave functions of the ring-shaped oscillator
potential which is given [29]
\begin{eqnarray}
V(r,\theta)=\kappa r^2+\frac{\omega}{r^2}+\frac{1}{r^2}\,\beta
cosec^2\theta\,,
\end{eqnarray}
Inserting this potential to the SE gives the following two
equations
\begin{subequations}
\begin{align}
\left\{\frac{d^2}{d\theta^2}-\frac{\xi+\beta M
}{sin^2\theta}+\ell^2\right\}
\Theta(\theta)&=0\,,\\
\left\{\frac{d^2}{dr^2}-M\kappa r^2
-\frac{L+M\omega}{r^2}+ME_{n\ell}\right\}R(r)&=0\,.
\end{align}
\end{subequations}
where the parameter $\ell$ is given as
$\ell=\sqrt{\beta+m^2\,}+\frac{1}{2}+s$ [15].

Defining a new variable $r^2=u$, taking a trial wave function as
$R=u^{-\tau/2}\phi(u)$ and using the following abbreviations
\begin{eqnarray}
\mu^2_{1}=-M\kappa\,\,;\,\,\mu^2_{2}=ME_{n\ell}\,\,;\,\,\tau(\tau+1)=L+M\omega
\end{eqnarray}
Eq. (41b) turns into
\begin{eqnarray}
u\frac{d^2\phi(u)}{du^2}-\left(\tau-\frac{1}{2}\right)\frac{d\phi(u)}{du}-\frac{1}{4}\,\left(\mu^2_{1}u-\mu^2_{2}
\right)\phi(u)=0\,,
\end{eqnarray}
Applying the Laplace transform to Eq. (43) we obtain a first-order
differential equation
\begin{eqnarray}
\left(t^2-\frac{\mu^2_{1}}{4}\right)\frac{df(t)}{dt}+\left\{\left(\tau+\frac{3}{2}\right)t-
\frac{\mu^2_{2}}{4}\right\}f(t)=0\,,
\end{eqnarray}
whose solution is
\begin{eqnarray}
f(t)=e'\left(t+\frac{\mu_{1}}{2}\right)^{-\frac{\mu^2_{2}}{4\mu_{1}}-\frac{1}{2}\left(\tau+\frac{3}{2}\right)}
\left(t-\frac{\mu_{1}}{2}\right)^{\frac{\mu^2_{2}}{4\mu_{1}}-\frac{1}{2}\left(\tau+\frac{3}{2}\right)}\,,
\end{eqnarray}
where $e'$ is a constant determining later.

We find a physically acceptable wave function (finite) only if
\begin{eqnarray}
\frac{\mu^2_{2}}{4\mu_{1}}-\frac{1}{2}\left(\tau+\frac{3}{2}\right)=n,\,\,n=0,
1, 2,\ldots
\end{eqnarray}
By using this requirement and expanding the function in Eq (45)
into series, we get
\begin{eqnarray}
f(t)=e''\sum_{k=0}^{n}\frac{(-1)^{k}n!}{(n-k)!k!}\,\left(t+\frac{\mu_{1}}{2}\right)^{-(\tau+\frac{3}{2}+k)}\,,
\end{eqnarray}
where $e''$ is another constant. Using the inverse Laplace
transform we obtain the solution of Eq. (43)
\begin{eqnarray}
\phi(u)=e'''e^{-\mu_{1}u/2}u^{\tau+1/2}\,_{1}F_{1}(-n,\tau+\frac{3}{2},u)\,,
\end{eqnarray}
where $e'''$ is a constant and used the property of the
hypergeometric functions given in Eq. (21). The radial wave
functions of the ring-shaped oscillator is
\begin{eqnarray}
R(u)=e_{n}e^{-\mu_{1}u/2}u^{(\tau+1)/2}\,_{1}F_{1}(-n,\tau+\frac{3}{2},u)\,,
\end{eqnarray}
where $e_{n}$ is normalization constant. Eq. (46) with the help of
Eq. (42) gives the energy spectrum of the ring-shaped
non-spherical oscillator potential
\begin{eqnarray}
E_{n\ell}=2\sqrt{\frac{\kappa}{M}\,}\left[2n+\sqrt{\ell^2+M\omega\,}+1\right]\,.
\end{eqnarray}

The ring-shaped oscillator potential [7] is obtained by setting
$\omega=0$ in Eq. (40) which gives from last equation
\begin{eqnarray}
E_{n\ell}=\sqrt{\frac{16\kappa}{M}\,}\left(n+\ell'+\frac{1}{2}\right)\,.
\end{eqnarray}
where $\ell'=\ell/2$ and last equation is the same result with the
ones obtained in Ref. [30] if setting $\kappa=\frac{\omega^2}{4}$
and $M=1$.

\section{Conclusions}
We have obtained the exact energy spectrum of some non-central
potentials such as Makarov potential, modified-Kratzer plus a
ring-shaped potential, double ring-shaped Kratzer potential,
modified non-central potential and ring-shaped non-spherical
oscillator potential by applying the Laplace transform approach to
the related part of the Schrödinger equation in spherical
coordinates. We have also obtained the corresponding
eigenfunctions of the above diatomic potentials. We have discussed
briefly some special cases of the potentials and observed that our
analytical results and also the results for the special cases are
the same with the ones obtained in literature. It is shown that
the Laplace transform approach is an applicable formalism to
obtain the energy spectrum and the eigenfunctions of some
non-central potentials.

\section{Acknowledgments}
One of the authors (A.A.) would like thank to Dr. O. Aydoðdu from
University of Mersin for his inexpressible help. This research was
partially supported by the Scientific and Technical Research
Council of Turkey.

\newpage

\begin{table}
\begin{ruledtabular}
\caption{Energy eigenvalues of the modified non-central potential
for different values of quantum numbers in $eV$ for $N_{2}$
molecule.}
\begin{tabular}{@{}cclccc@{}}
&&&\multicolumn{2}{c}{$E_{n\ell}$} & Ref. [26] \\
\cline{4-5}
$n$ & $s$  & $m$ & $\beta=\gamma=0.1$ & $\beta=\gamma=0$ &    \\
0 & 0 & 0 & 11.93837780671 & 0.05443703 & 0.054430 \\
1 & 0 & 0 & 11.93837780698 & 0.16207785 & 0.162057 \\
  & 1 & 1 & 11.93837780740 & 0.16354346 & 0.162546 \\
2 & 0 & 0 & 11.93837780726 & 0.26826281 & 0.268229 \\
  & 1 & 1 & 11.93837778077 & 0.26970864 & 0.268711 \\
  & 2 & 2 & 11.93837780810 & 0.27308086 & 0.269675 \\
3 & 0 & 0 & 11.93837780754 & 0.37301804 & 0.372972  \\
  & 1 & 1 & 11.93837780800 & 0.37444445 & 0.373447 \\
  & 2 & 2 & 11.93837780840 & 0.37777137 & 0.374398 \\
  & 3 & 3 & 11.93837780880 & 0.38299550 & 0.375823 \\
\end{tabular}
\end{ruledtabular}
\end{table}

\end{document}